\documentclass[12pt,epsf]{article}
 \usepackage[compat=1.1.0]{tikz-feynman}
 \usepackage{verbatim,graphics,graphicx,color,slashed,amsmath,subcaption}
 \usepackage{bm}
 \usepackage{cite}
 \usepackage{amssymb}
 \usepackage{braket}
 \usepackage{ascmac}
 \usepackage{multirow}
 \usepackage[compat=1.1.0]{tikz-feynman} 
\usepackage[]{hyperref}
 \hypersetup{colorlinks,bookmarks,unicode,linktocpage=true,linkcolor=blue, anchorcolor=blue, citecolor=blue}
 \usepackage{comment} 
\newcounter{num}

\begin{document}
\thispagestyle{empty}
\vspace*{-15mm}
\baselineskip 1pt
\begin{flushright}
\begin{tabular}{l}
\end{tabular}
\end{flushright}
\baselineskip 24pt
\vglue 10mm

\vspace{15mm}
\begin{center}
{\Large\bf Higgs-Portal Spin-1 Dark Matter with
Parity-Violating Interaction for a Galactic Halo Gamma Ray Excess}
\vspace{7mm}

\baselineskip 18pt
{\bf Kimiko Yamashita}
\vspace{2mm}

{\it 
\, \, \, \, Department of Physics, Ibaraki University, Mito 310-8512, Japan
\newline \newline
kimiko.yamashita.nd93@vc.ibaraki.ac.jp}\\
\vspace{10mm}
\end{center}
\begin{center}
\begin{minipage}{14cm}
\baselineskip 16pt
\noindent

\end{minipage}
\end{center}

\baselineskip 18pt
\def\thefootnote{\fnsymbol{footnote}}
\setcounter{footnote}{0}
\begin{abstract}
We study a dark photon dark matter scenario associated with a gauged $U(1)_X$ symmetry,
stabilized by a dark parity that forbids kinetic mixing with the Standard Model.
The leading interactions between the dark photon dark matter and the Standard Model
arise from dimension-six Higgs-portal operators.
In previous work, we found that for the parity-violating operator,
the dark matter annihilation process is $p$-wave suppressed,
naturally evading stringent direct-detection constraints
while reproducing the observed relic abundance through thermal freeze-out.
For a cutoff scale of $1~\mathrm{TeV}$, a dark matter mass of around $400~\mathrm{GeV}$
is favored to realize the observed relic abundance.
This scenario predicts cosmic-ray signals, in particular in the $W^{+}W^{-}$ channel,
which can be targeted by indirect-detection experiments.
Recently, a halo-like gamma-ray excess has been reported by Totani.
Assuming a Navarro--Frenk--White-$\rho^2$ morphology for the dark matter distribution,
a dark matter mass of $420~\mathrm{GeV}$ is favored for the $W^{+}W^{-}$ final state.
Motivated by this excess, we consider the present dark photon dark matter framework
augmented by a CP-even sub-GeV scalar mediator,
which couples to the dark photon mass operator in order to enhance the
present-day annihilation rate in the Galactic halo while inducing only an $\mathcal{O}(1)$ correction at freeze-out.
The exchange of this scalar induces a long-range attractive force,
leading to Sommerfeld enhancement of the annihilation rate.
The enhancement is saturating in dwarf spheroidal galaxies and the annihilation rate is dropping as $p$-wave remaining consistent with constraints from dwarf galaxies and cosmology.
\end{abstract}
\newpage
\section{Introduction}\label{sec:introduction}

Dark photons, defined as gauge bosons of an extra $U(1)$ gauge symmetry,
provide a simple extension of the Standard Model (SM)
and have been studied extensively in a wide range of contexts.
In the conventional scenario, the dark photon interacts with the SM
through kinetic mixing with the $U(1)_Y$ gauge boson~\cite{Holdom:1985ag}.
At energies well below the electroweak scale,
this interaction can be effectively interpreted
as a mixing between the dark photon and the SM photon
after integrating out the $Z$ boson~\cite{Fabbrichesi:2020wbt}.
Experimental and observational constraints severely restrict
the kinetic mixing parameter, typically to the range
$10^{-16}$--$10^{-5}$ depending on the dark photon mass,
over a wide mass range of $10^{-17}$--$10^{5}$~eV,
well below the $Z$-boson mass~\cite{Caputo:2021eaa}.
For TeV-scale dark photons, collider constraints become weaker
due to the strong suppression of production cross sections
by small mixing angles, e.g., for a mixing angle of 0.01,
the dark photon production cross section at the $13$~TeV Large Hadron Collider
is expected to be well below the order of fb~\cite{Yamashita:2024krp}.

Nevertheless, the requirement of tiny/relatively small kinetic mixing
motivates alternative scenarios in which the mixing is forbidden
by symmetry.
In particular, if the dark photon is odd under a dark parity,
whereas all SM particles are even,
kinetic mixing is absent and the dark photon is stable,
making it a viable dark matter (DM) candidate.
In such frameworks, the leading interactions between the dark photon
and the SM sector arise from higher-dimensional Higgs-portal operators.

In Ref.~\cite{Yamashita:2024krp},
we studied dimension-six Higgs-portal operators for dark photon DM.
It was shown that the parity-violating operator naturally evades stringent
direct-detection constraints, since the tree-level
DM--nucleon scattering amplitude vanishes in the zero momentum transfer limit.
As a result, the observed DM relic abundance
can be successfully reproduced through thermal freeze-out.
The same parity-violating structure also implies that
the DM annihilation cross section is $p$-wave suppressed.
For a cutoff scale of $1~\mathrm{TeV}$, a DM mass of around $400~\mathrm{GeV}$
is favored to realize the observed relic abundance.

This scenario predicts potentially observable indirect-detection
signals, in particular in the $W^{+}W^{-}$ channel.

Recently, Totani analyzed 15 years of \textit{Fermi} Large Area Telescope data
and reported a halo-like excess
in the Galactic diffuse gamma-ray emission,
with a spectral peak around $E_\gamma\simeq20~\mathrm{GeV}$~\cite{Totani:2025fxx}.
The excess can be fitted by annihilation of DM
with a mass $m_X=420$~GeV
and a thermally averaged annihilation cross section
$\langle\sigma v\rangle_{\mathrm{halo}}\sim10^{-24}\,\mathrm{cm}^3\,\mathrm{s}^{-1}$
assuming a Navarro--Frenk--White (NFW)-$\rho^2$ DM morphology
for the $W^{+}W^{-}$ final state~\cite{Totani:2025fxx}.

The outstanding puzzle is that this cross section exceeds both the
canonical thermal relic value,
$\langle \sigma v \rangle_{\rm fo} \sim 10^{-26}~\mathrm{cm}^3\,\mathrm{s}^{-1}$,
and the limits from dwarf spheroidal galaxies,
$\langle \sigma v \rangle_{\rm dwarf} < 10^{-25}~\mathrm{cm}^3\,\mathrm{s}^{-1}$.
Several theoretical mechanisms have been proposed to reconcile this
tension by exploiting the strong hierarchy of DM velocities among
thermal freeze-out, $v_{\rm fo}\sim0.3$, the Galactic halo,
$v_{\rm halo}\sim10^{-3}$, and dwarf spheroidal galaxies,
$v_{\rm dwarf}\sim10^{-5}$--$10^{-4}$~\cite{Murayama:2025ihg,Jho:2025iah,Nomura:2026ntp,Yoshimatsu:2026gff}.

In this work, we extend the parity-violating Higgs-portal framework
of Ref.~\cite{Yamashita:2024krp}
by introducing a light CP-even scalar field
that couples directly to the dark photon mass operator.
This coupling induces a long-range attractive interaction
between DM particles,
leading to Sommerfeld enhancement~\cite{Sommerfeld:1931qaf,Hisano:2003ec,Hisano:2002fk} of the annihilation rate.
The parity-violating Higgs-portal interaction
continues to determine the short-distance annihilation process,
ensuring $p$-wave suppression,
while the scalar-mediated force generates
a velocity-dependent enhancement
that becomes effective in the Galactic halo
but saturates in low-velocity environments.
As a result, the observed relic abundance,
the Galactic halo gamma-ray excess,
and constraints from dwarf spheroidal galaxies and the cosmic microwave background (CMB)~\cite{Padmanabhan:2005es,Slatyer:2015jla,Planck:2018vyg}
can be simultaneously accommodated.

The idea of combining $p$-wave-suppressed annihilation with Sommerfeld
enhancement has also been explored in Ref.~\cite{Jho:2025iah}.
A virtue of the present Higgs-portal realization is that it naturally
evades direct-detection constraints, since we do not introduce any mixing
between the light scalar mediator and the SM Higgs boson.

This paper is organized as follows.
In Sec.~\ref{sec:eft}, we introduce the effective field theory framework
for the Higgs-portal spin-1 DM model and the scalar-mediated long-range interaction.
In Sec.~\ref{sec:pheno}, we discuss the DM relic abundance
and provide an interpretation of the gamma-ray signal
from the Galactic halo, including constraints from
dwarf spheroidal galaxies and the CMB.
Finally, we summarize our conclusions in Sec.~\ref{sec:summary}.

\section{Higgs-Portal Spin-1 Dark Matter for Sommerfeld Enhancement}
\label{sec:eft}

We consider the dark photon $X_\mu$ as the gauge boson of an extra
$U(1)_X$ gauge symmetry.
The dark photon is assumed to be odd under a dark parity,
whereas all SM particles are neutral under $U(1)_X$
and even under the dark parity.
As a result, kinetic mixing between the dark photon
and the SM hypercharge gauge boson is forbidden,
and the dark photon is stable,
serving as a viable DM candidate.

The leading couplings arise from the following dimension-six Higgs-portal operators.
\begin{align}
\mathcal{O}
&=
(H^\dagger H)\,X_{\mu\nu}X^{\mu\nu},
\\
\tilde{\mathcal{O}}
&=
(H^\dagger H)\,X_{\mu\nu}\tilde X^{\mu\nu},
\end{align}
where $H$ is the SM Higgs doublet and
$X_{\mu\nu}=\partial_\mu X_\nu-\partial_\nu X_\mu$
is the field strength of the vector DM.
The dual field strength is defined as
$\tilde X_{\mu\nu}=1/2\epsilon_{\mu\nu\rho\sigma}X^{\rho\sigma}$.

The Higgs-portal interaction is written as
\begin{align}
\mathcal{L}_{\rm Higgs\text{-}portal}
=
\frac{C}{\Lambda^2}\,\mathcal{O}
+
\frac{\tilde C}{\Lambda^2}\,\tilde{\mathcal{O}},
\end{align}
where $C$ and $\tilde C$ are Wilson coefficients and $\Lambda$ is the characteristic scale of new physics.
The parity-odd operator $\tilde{\mathcal{O}}$
leads to $p$-wave-suppressed annihilation
and vanishing tree-level direct-detection amplitudes
in the zero momentum-transfer limit~\cite{Yamashita:2024krp}.
In this work, we concentrate on the parity-odd operator, $\tilde{\mathcal{O}}$.

To generate long-range interactions,
we introduce a light CP-even scalar field $\phi$
that couples directly to the dark photon mass operator $X_\mu X^\mu$,
\begin{align}
\mathcal{L}_{\rm SE}
=
\frac{1}{2}\,\mu\,\phi\,X_\mu X^\mu,
\label{eq:lag_SE}
\end{align}
where $\mu$ is a dimensionful coupling.
Such an interaction softly breaks the $U(1)_X$ gauge symmetry
at low energies and can arise in effective theories
below the scale of $U(1)_X$ symmetry breaking.

The exchange of the scalar $\phi$
between non-relativistic DM particles
induces an attractive Yukawa potential,
leading to Sommerfeld enhancement of the annihilation rate.

In the present work, we do not assume sizable direct couplings between $\phi$ and SM particles. Consequently, laboratory and cosmological constraints on the mediator are model dependent and rely on the details of the ultraviolet (UV) completion. 
A dedicated analysis of such constraints is beyond the scope of the present work and is left for future investigation.

Note that perturbative unitarity is not violated at high energies, since
the $X_L X_L \to X_L X_L$ scattering amplitude via scalar exchange
exhibits a cancellation among the $s$-, $t$-, and $u$-channel contributions
at order $\mu^2 p^2/m_X^4$.
As a result, the total amplitude without angular dependence is given by $-\mu^2/s$.
Here, $X_L$ denotes the longitudinally polarized mode of the DM particle,
$p$ is the momentum of the DM particles in the center-of-mass frame,
and $s$ is the Mandelstam variable.


The total interaction relevant for DM annihilation is 
\begin{align}
\mathcal{L}_{\rm int}
=
\frac{\tilde C}{\Lambda^2}
(H^\dagger H)\,X_{\mu\nu}\tilde X^{\mu\nu}
+
\frac{1}{2}\,\mu\,\phi\,X_\mu X^\mu.
\label{eq:3-point}
\end{align}

In a possible UV completion,
the dark photon mass and the three-point interaction in Eq.~\eqref{eq:3-point}
may arise from a Higgsed $U(1)_X$ gauge theory
or through a Stueckelberg mechanism~\cite{DeFelice:2025qva}.
It is worth noting that, although the Stueckelberg mechanism provides
a ``gauge-invariant'' mass and interaction terms proportional to
$X^\mu X_\mu$ for the vector boson without introducing
a physical Higgs field at low energies,
it can be regarded as an effective description of
a spontaneously broken (Higgsed) gauge theory
in the ultraviolet completion.
Since our analysis is performed within an effective field theory framework,
we remain agnostic about the microscopic origin
of the vector mass and interaction.

Note that the benchmark point corresponds to $\mu \simeq 940~{\rm GeV}$, $\Lambda = 1~{\rm TeV}$, and a DM mass of $420~{\rm GeV}$. Therefore, the present benchmark should be regarded as lying near the boundary of the effective field theory validity regime. The cutoff scale $\Lambda$ characterizes the Higgs-portal effective operator, whereas the scalar-mediated interaction responsible for Sommerfeld enhancement is treated as an additional low-energy interaction. For $m_X=420~{\rm GeV}$, higher-dimensional operators beyond the dimension-six level may induce corrections at the $\mathcal{O}(20\%)$ level, as $(m_X/\Lambda)^2 \simeq 0.18$.
Nevertheless, the benchmark remains useful for illustrating the phenomenology of the Sommerfeld-enhanced Higgs-portal scenario.
The purpose of this work is to provide a phenomenological investigation of the Sommerfeld-enhanced Higgs-portal scenario, and a UV-complete realization is left for future work.

We also note that we can translate the mass of a new heavy resonance, $M \geq 1$~TeV, 
into the Wilson coefficients of the dimension-six operators~\cite{Boughezal:2022nof,Yamashita:2024krp} as 
\begin{align} 
\frac{\Lambda}{\sqrt{|C_6|}} \geq \frac{1~\mathrm{TeV}}{g}, 
\end{align} 
where $C_6$ is the Wilson coefficient of a dimension-six operator, and $g$ is the coupling of the heavy resonance. 
Then, if we take $g=\sqrt{4\pi}$ as its maximum value and $\Lambda = 1$~TeV, we have $|C_6| \leq 4\pi$. 
Thus, it should be an acceptable DM effective field theory scenario with $|\tilde C|=1$ and $\Lambda = 1$~TeV.


\section{Dark Matter Relic Abundance and a Galactic Halo Gamma Ray Excess}
\label{sec:pheno}

In this section, we study the relic abundance of Higgs-portal spin-1 DM
and its indirect-detection signatures in the Galactic halo.
We assume that the present-day DM abundance
is determined by the standard thermal freeze-out mechanism.

\begin{figure}[!t]
\begin{center}
	\begin{tikzpicture}[baseline=($(int)$)]
		\begin{feynman}[inline=($(int)$),medium]
			\vertex (in1) at (-1.8, 1.8) {\( X \)};
			\vertex (in2) at (-1.8, -1.8) {\( X \)};
			\vertex (int) at (0, 0);
			\vertex (out1) at (1.8, 1.8) {\( h \)};
			\vertex (out2) at (1.8, -1.8) {\( h \)};
			\diagram* {
				(in1) -- [photon] (int) -- [scalar] (out2),
				(in2) -- [photon] (int) -- [scalar] (out1),
			};
		\end{feynman}
	\end{tikzpicture}
	\begin{tikzpicture}[baseline=($(int)$)]
		\begin{feynman}[inline=($(int)$),medium]
			\vertex (in1) at (-1.8, 1.8) {\( X \)};
			\vertex (in2) at (-1.8, -1.8) {\( X \)};
			\vertex (int) at (0, 0);
			\vertex (out) at (1.8, 0);
			\vertex (out1) at (3.6, 1.8) {\( \mathrm{SM}(V,h,f) \)};
			\vertex (out2) at (3.6, -1.8) {\( \mathrm{SM}(V,h,f) \)};
			\diagram* {
				(in1) -- [photon] (int),
				(in2) -- [photon] (int),
				(int) -- [scalar, edge label'=\(h\)] (out),
				(out) -- [scalar] (out1),
				(out) -- [scalar] (out2),
			};
		\end{feynman}
	\end{tikzpicture}
\end{center}
\caption{Feynman diagrams for DM annihilation induced by the Higgs-portal interaction.}
\label{fig:diagram_RelicD}
\vspace{3mm}
\end{figure}

The relic density of Higgs-portal vector DM with parity-violating
dimension-six interactions was studied in detail in Ref.~\cite{Yamashita:2024krp}.
The dominant annihilation channels,
$XX\to hh$, $W^+W^-$, $ZZ$, and $f\bar f$,
are induced by the Higgs-portal operator
$(H^\dagger H)X_{\mu\nu}\tilde X^{\mu\nu}$ (Figure~\ref{fig:diagram_RelicD}).
Here, $h$ denotes the SM Higgs boson,
$V = W^\pm, Z$, and $f$ represents the SM fermions.
The parity-odd structure leads to $p$-wave suppression,
$\sigma v \propto v^2$, and simultaneously causes the tree-level
DM--nucleon scattering amplitude to vanish at zero momentum transfer that
allows the model to evade stringent direct-detection
constraints. Here, $v$ denotes the relative velocity of the DM particles.

For DM masses in the few-hundred-GeV range and a cutoff scale
$\Lambda\sim\mathcal{O}(1~\mathrm{TeV})$,
the observed relic abundance,
$\Omega_{\rm DM}h^2\simeq0.12$,
is reproduced through thermal freeze-out,
corresponding to a typical thermally averaged annihilation cross section
\[
\langle\sigma v\rangle_{\rm fo}\sim10^{-26}~\mathrm{cm}^3\,\mathrm{s}^{-1}.
\]

\begin{figure}[t]
\centering
\includegraphics[width=90mm]{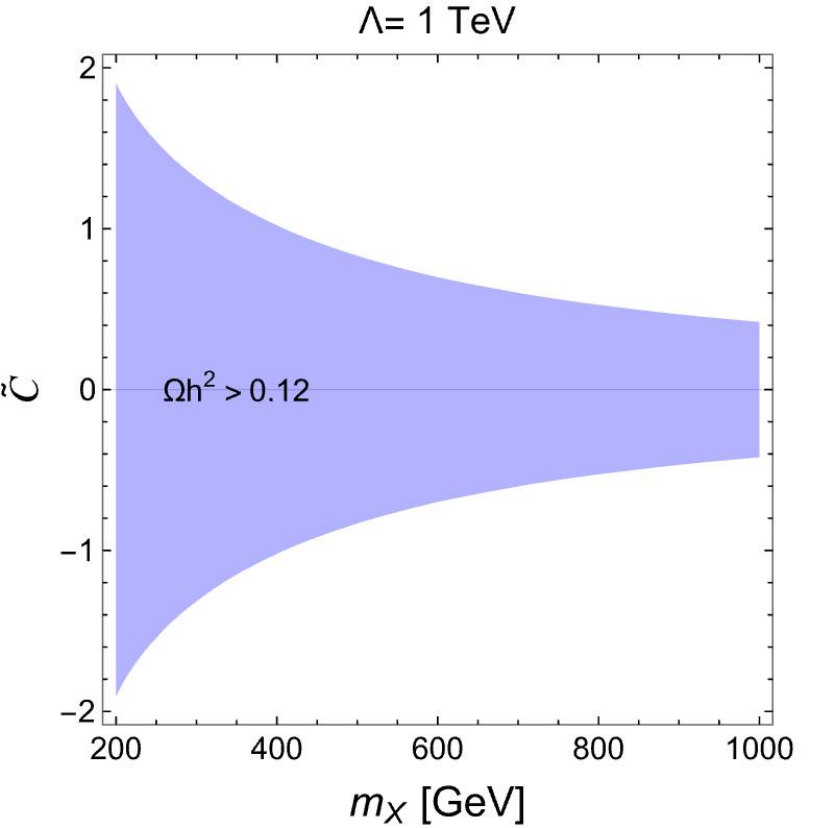}
\caption{
Relic abundance in the parameter space of $(m_X, \tilde{C})$ for $\Lambda = 1~\mathrm{TeV}$.
The relic abundance of DM is overproduced in the blue regions,
corresponding to $\Omega_{\rm DM} h^2 > 0.12$, and saturates the observed
value along the boundary of the blue region.
The benchmark value $m_X = 420~\mathrm{GeV}$ is favored for $|\tilde{C}| \sim 1$.
}
\label{fig:Ctilde_mX}
\end{figure}

Figure~\ref{fig:Ctilde_mX} shows the relic abundance in the parameter space
of $(m_X, \tilde{C})$ for $\Lambda = 1~\mathrm{TeV}$.
For $m_X\simeq420$~GeV, used below as a benchmark,
we find $|\tilde{C}| \sim 1$.
This mass is directly motivated by the value required to explain
the Galactic halo gamma-ray excess reported by Totani, i.e., $m_X = 420$~GeV for DM annihilation dominantly into the $W^+W^-$ final state.

In fact, for a DM mass of $420$~GeV,
the annihilation in this model is dominated by the $W^+W^-$ channel,
followed by $hh$ and $ZZ$, with the $hh$ and $ZZ$ rates being roughly half that of $W^+W^-$, respectively.
Accordingly, we focus on the $W^{+}W^{-}$ channel in our analysis of the
Galactic halo gamma-ray excess.
Including the $hh$ and $ZZ$ channels changes the total annihilation rate only by an $\mathcal{O}(1)$ factor and does not qualitatively affect the conclusions presented below.

We emphasize that the light CP-even scalar field $\phi$ introduced in Sec.~\ref{sec:eft} to mediate long-range interactions gives only an $\mathcal{O}(1)$ correction to the freeze-out annihilation rate, in contrast to the much larger Sommerfeld enhancement in present-day low-velocity environments.
Sommerfeld enhancement becomes operative particularly when the de Broglie
wavelength of DM exceeds the range of the Yukawa potential, i.e., when $v \lesssim m_\phi/m_X$~\cite{ArkaniHamed:2008qn,Cassel:2009wt}~\footnote{depending also on the coupling strength in the Yukawa potential, $\alpha_{\rm eff}(m_X, \mu)$}.
Since $v_{\rm fo} \sim 0.3 \gg m_\phi/m_X$ for a DM mass
$m_X = 420~\mathrm{GeV}$ and a mediator mass of
$\mathcal{O}(100)~\mathrm{MeV}$, as considered below, 
Sommerfeld enhancement at freeze-out is much less significant than in present-day low-velocity environments. 
Here, $v_{\rm fo}$ denotes the typical
relative velocity of DM particles at thermal freeze-out.

In the Milky Way halo, DM is highly non-relativistic, with a typical velocity
$v_{\rm halo} \sim 10^{-3}$.
In this low-velocity environment, the attractive force mediated by the scalar induces a
Yukawa potential,
leading to Sommerfeld enhancement of the annihilation rate through
an increased probability at short distances relevant for annihilation.

In the present model, $\phi$ couples to the dark photon mass operator via
$\frac{1}{2}\mu\,\phi\,X_\mu X^\mu$,
inducing an attractive Yukawa potential
\begin{align}
V_Y(r)&=-\alpha_{\rm eff}\frac{e^{-m_\phi r}}{r}, \label{eq:Yukawa-potential}\\
\alpha_{\rm eff}&\equiv\frac{\mu^2}{16\pi m_X^2}. \label{eq:alpha_eff}
\end{align}
Sommerfeld enhancement grows as the velocity decreases
until it saturates at $v\lesssim m_\phi/m_X$.

For $p$-wave annihilation in the Coulomb-like regime, and in the low-velocity limit,
the Sommerfeld enhancement factor behaves as
$S_1(v) \simeq \pi \alpha_{\rm eff}^3 / v^3$, up to finite-range effects.
As a result, for $\alpha_{\rm eff}=0.2$ and a mediator mass
$m_\phi = 400~\mathrm{MeV}$, the present-day annihilation rate in the
Galactic halo can reach
\[
\langle \sigma v \rangle_{\rm halo}
\sim
10^{-24}~\mathrm{cm}^3\,\mathrm{s}^{-1},
\]
which is consistent with the observed gamma-ray excess.
From Eq.~\eqref{eq:alpha_eff}, a dimensionful coupling is given by $\mu \simeq 1.3~\mathrm{TeV}$ for
a DM mass of $m_X = 420~\mathrm{GeV}$ and $\alpha_{\rm eff}=0.2$.

For illustration, we first consider $\alpha_{\rm eff}=0.2$ 
(corresponding to $\mu \simeq 1.3~\mathrm{TeV}$ for $m_X = 420~\mathrm{GeV}$) 
and $m_\phi = 400~\mathrm{MeV}$.

For $p$-wave annihilation ($\ell=1$), the Sommerfeld enhancement factor
in the analytic approximation of Ref.~\cite{Cassel:2009wt} is given by
\begin{align}
S_1(v)
&=
\left|
\dfrac{\Gamma(a^-)\Gamma(a^+)}{\Gamma(2+2iw)}
\right|^2,
\label{eq:approx}
\end{align}
where
\begin{align}
\delta \equiv  \frac{\pi^2}{6}\,m_\phi,
\qquad
w \equiv \frac{m_X v}{2\delta},
\qquad
x \equiv \frac{2\alpha_{\rm eff}}{v}, \\
a^{\pm}
=
2+i w
\left(
1 \pm \sqrt{1-\frac{x}{w}}
\right).
\end{align}
This analytic expression is obtained by approximating the Yukawa potential
with the Hulth\'en potential and by employing a modified centrifugal term, i.e., a Manning--Rosen or Eckart potential,
which allows for an analytic solution~\cite{Cassel:2009wt}~\footnote{For the Coulomb limit, 
$S_1(v) = S_0 (1+\alpha^2_\mathrm{eff}/v^2)$, where $S_0 = 2\pi \alpha_\mathrm{eff}/\{v[1-\exp(-2\pi\alpha_\mathrm{eff}/v)]\}$~\cite{Cassel:2009wt}.}.

\begin{figure}[t]
  \centering
  \begin{minipage}[t]{0.48\textwidth}
    \centering
    \includegraphics[width=\textwidth]{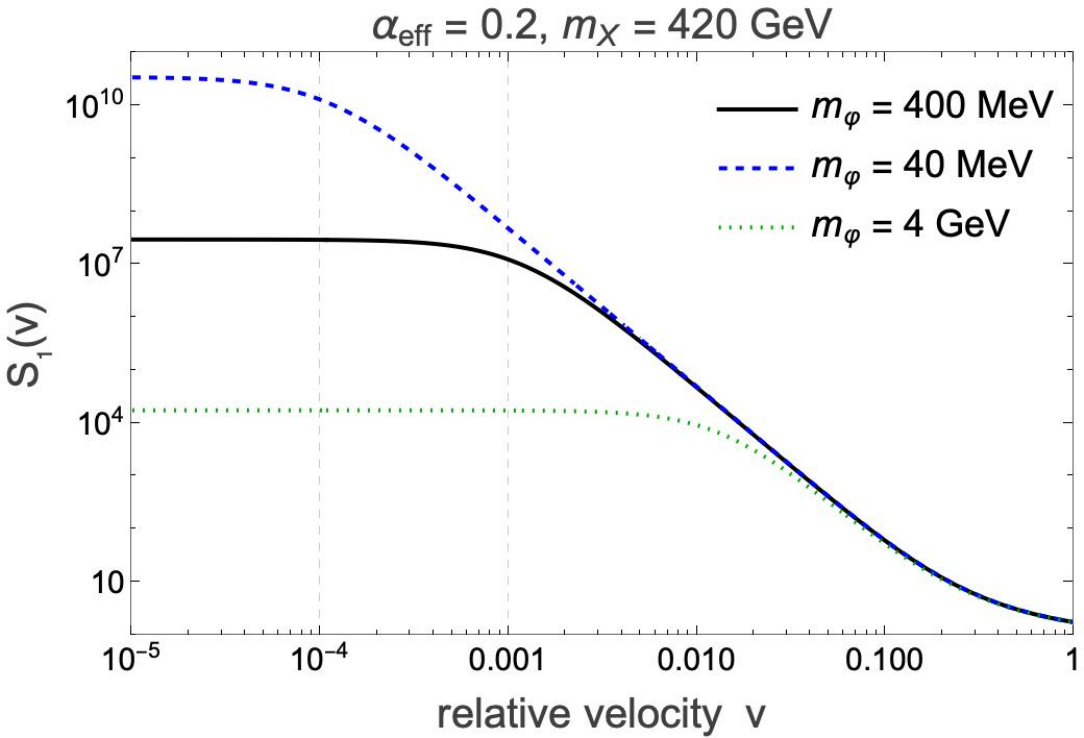}
    \subcaption{Sommerfeld enhancement factor $S_1(v)$.}
    \label{fig:S1}
  \end{minipage}
  \hfill
  \begin{minipage}[t]{0.48\textwidth}
    \centering
    \includegraphics[width=\textwidth]{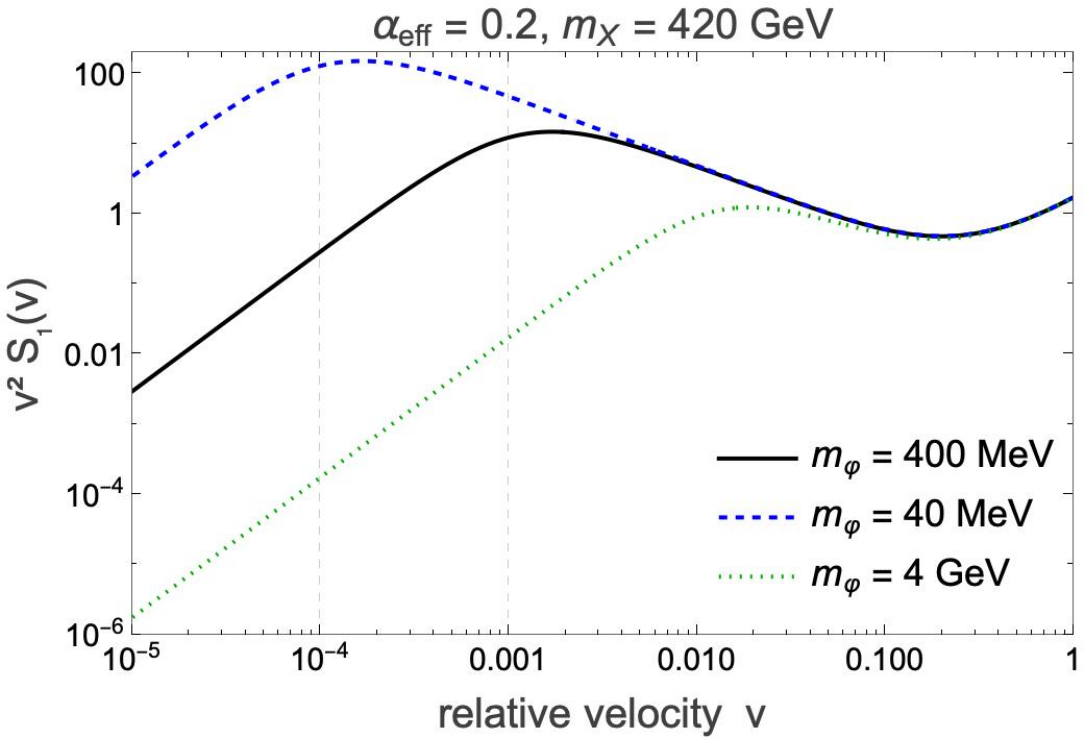}
    \subcaption{Effective annihilation rate $v^2 S_1(v)$.}
    \label{fig:sigmaEff}
  \end{minipage}
  \caption{
Velocity dependence of Sommerfeld-enhanced $p$-wave annihilation
for $m_X = 420~\mathrm{GeV}$ and $\alpha_{\rm eff} = 0.2$.
The mediator mass is varied as
$m_\phi = 400~\mathrm{MeV}$ (solid black),
$40~\mathrm{MeV}$ (dashed blue),
and $4~\mathrm{GeV}$ (dotted green).
The left panel shows the Sommerfeld enhancement factor $S_1(v)$,
while the right panel displays the effective annihilation rate
$v^2 S_1(v)$.
The Sommerfeld factor is evaluated using the analytic approximation
of Cassel~\cite{Cassel:2009wt}, Eq.~\eqref{eq:approx}.
  }
  \label{fig:SE_pair}
\end{figure}
Figure~\ref{fig:SE_pair} shows the interplay between Sommerfeld enhancement
and intrinsic $p$-wave suppression using the analytic approximation in Eq.~\eqref{eq:approx}.
At large velocities, the Sommerfeld factor approaches unity.
As the velocity decreases to values relevant for the Galactic halo,
the enhancement grows rapidly,
while finite-range effects of the Yukawa interaction lead to saturation
at very small velocities.

As a representative benchmark, we adopt $m_X=420~{\rm GeV}$, which is motivated by the DM interpretation of the Galactic halo gamma-ray excess reported by Totani~\cite{Totani:2025fxx}. The qualitative dependence of the Sommerfeld enhancement on the model parameters is illustrated in Figure~\ref{fig:SE_pair}, where results for three different mediator masses are shown.

As a result, for a light scalar mass of $m_{\phi} = 400$~MeV, the effective annihilation rate $v^2 S_1(v)$
is maximized near $v_\mathrm{halo}\sim10^{-3}$
and becomes strongly suppressed in dwarf spheroidal galaxies,
where $v_{\rm dwarf}\sim10^{-5}$--$10^{-4}$.
Consequently, the model can simultaneously account for the Galactic halo
gamma-ray excess while remaining consistent with
dwarf-galaxy and cosmological constraints.

The result is exact in the Coulomb limit and provides a controlled
approximation away from resonance regions~\cite{Cassel:2009wt}, where resonant states can lead
to additional Sommerfeld enhancement~\cite{ArkaniHamed:2008qn,Cassel:2009wt}.
Such resonances occur only within a limited velocity range and therefore
do not explicitly appear in the plots.
They can be smeared out by velocity averaging, and are
thus not considered in this work.

\begin{figure}[t]
  \centering
  \begin{minipage}[t]{0.47\textwidth}
    \centering
    \includegraphics[height=6.5cm, keepaspectratio]{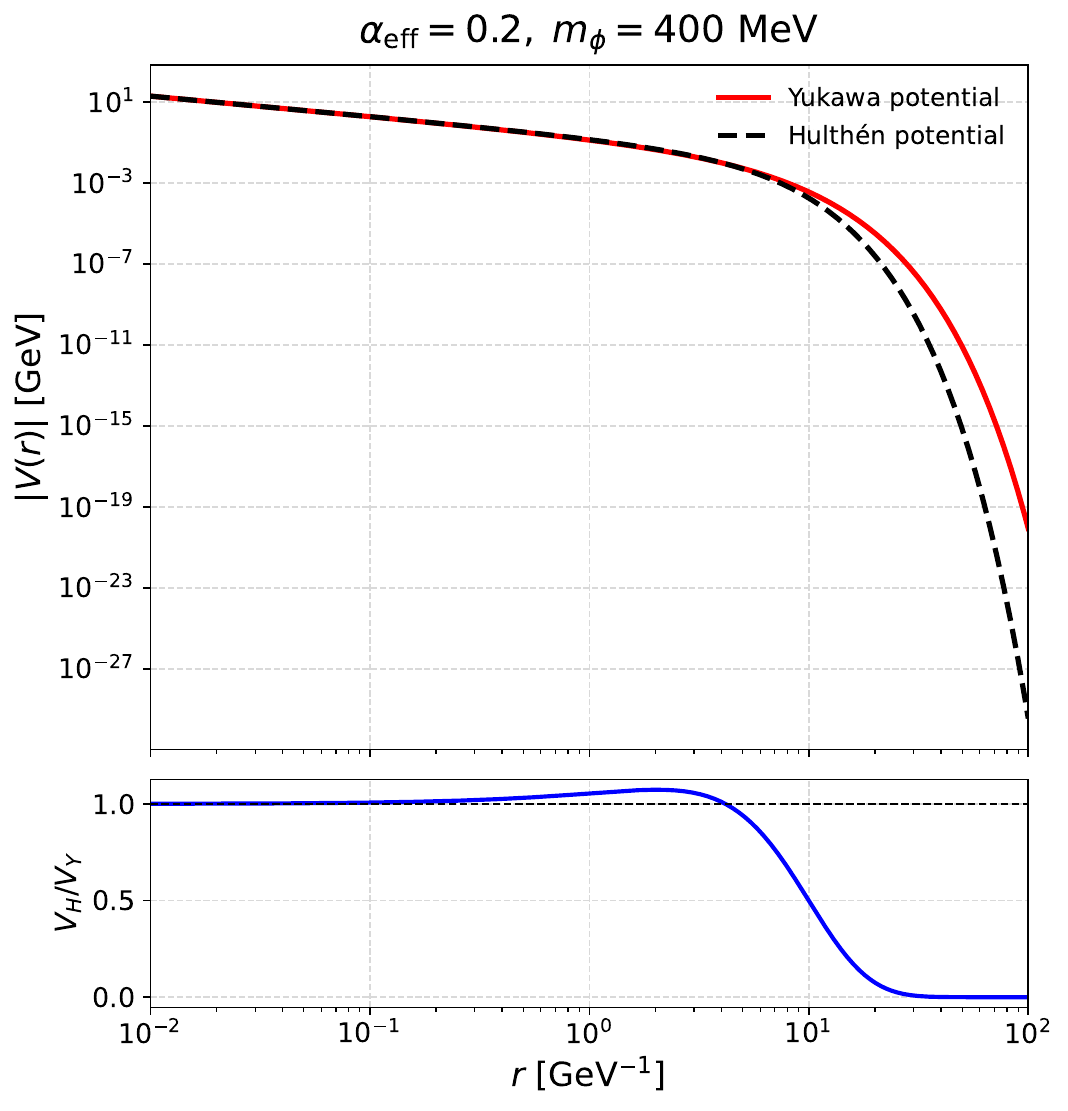}
    \subcaption{Comparison between the Yukawa potential (solid red) and the Hulth\'en potential (dashed black),
	with the lower panel showing the ratio $V_{\rm Hulth\acute{e}n}/V_{\rm Yukawa}$}
    \label{fig:potential}
  \end{minipage}
  \hspace{0.03\textwidth}
    \begin{minipage}[t]{0.47\textwidth}
    \centering
    \includegraphics[height=6.5cm, keepaspectratio]{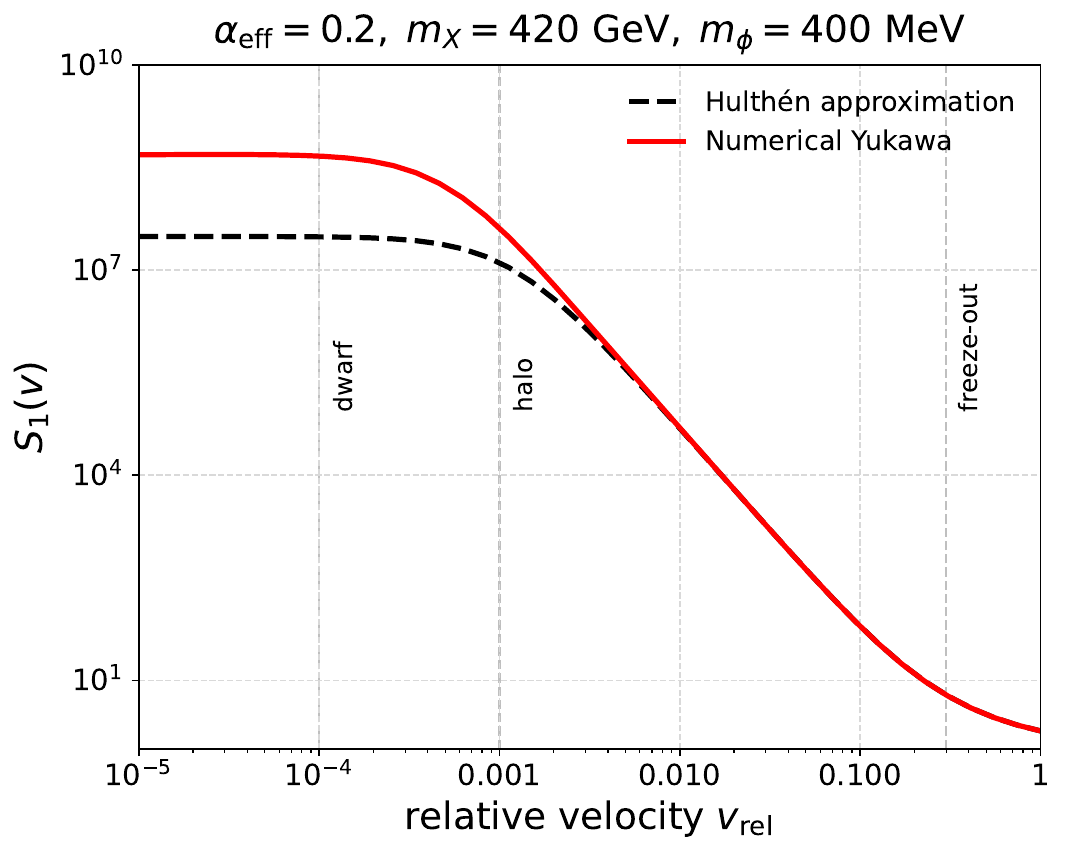}
\subcaption{Comparison between the numerical solution of the Schr\"odinger equation with the Yukawa potential (solid red) and the analytic Hulth\'en approximation (dashed black) for $m_X = 420~\mathrm{GeV}$}
    \label{fig:SE_comparison}
  \end{minipage}
  
  \caption{Both figures use $\alpha_{\rm eff}=0.2$ and $m_\phi = 400~\mathrm{MeV}$.}
\end{figure}
We now compare the numerical solution for the exact Yukawa potential with the analytic  Hulth\'en approximation of Ref.~\cite{Cassel:2009wt}, which we have used throughout the preceding discussion.
The result is shown in Figure~\ref{fig:SE_comparison}.
For deriving the numerical solution, we follow the method of Ref.~\cite{Iengo:2009ni}.
The exact Yukawa potential is given by 
\begin{align} 
V_Y(r) = -\alpha_{\rm eff} \frac{e^{-m_\phi r}}{r}, \tag{\ref{eq:Yukawa-potential}}
\end{align} 
while the Hulth\'en approximation employed in Ref.~\cite{Cassel:2009wt} is 
\begin{align} 
V_H(r) = -\alpha_{\rm eff} \frac{\delta\,e^{-\delta r}} {1-e^{-\delta r}}, \qquad \delta=\frac{\pi^2}{6}m_\phi. \end{align}
The corresponding potentials are shown in Figure~\ref{fig:potential}.

The numerical Yukawa solution exhibits a clear saturation behavior at sufficiently low velocities and does not show the sharp resonant structure expected for a narrow Sommerfeld resonance. We therefore conclude that the benchmark point is not located on a narrow Sommerfeld resonance.

We find good agreement between the two approaches in the velocity range relevant for thermal freeze-out. At very low velocities relevant for dwarf spheroidal galaxies, however, the numerical Yukawa solution predicts a larger Sommerfeld enhancement factor than the Hulth\'en approximation, with the difference reaching approximately one order of magnitude.

This behavior can be understood from Figure~\ref{fig:potential}. While the two potentials agree well at short distances, their difference gradually increases at larger distances. Since Sommerfeld enhancement is particularly sensitive to the long-range behavior of the potential at low velocities, such deviations can lead to noticeable differences between the numerical Yukawa solution and the analytic Hulth\'en approximation in the dwarf-galaxy velocity regime~\footnote{We note that the modified centrifugal term employed in the Manning--Rosen or Eckart potential does not reproduce the exact long-distance behavior of the centrifugal term~\cite{Cassel:2009wt} and may also contribute to the difference between the analytic and numerical results at low velocities.}.

In the following analysis, we therefore use the numerical Yukawa results for the final benchmark predictions and the evaluation of indirect-detection constraints.
\begin{figure}[t]
  \centering
  \includegraphics[width=0.6\textwidth]{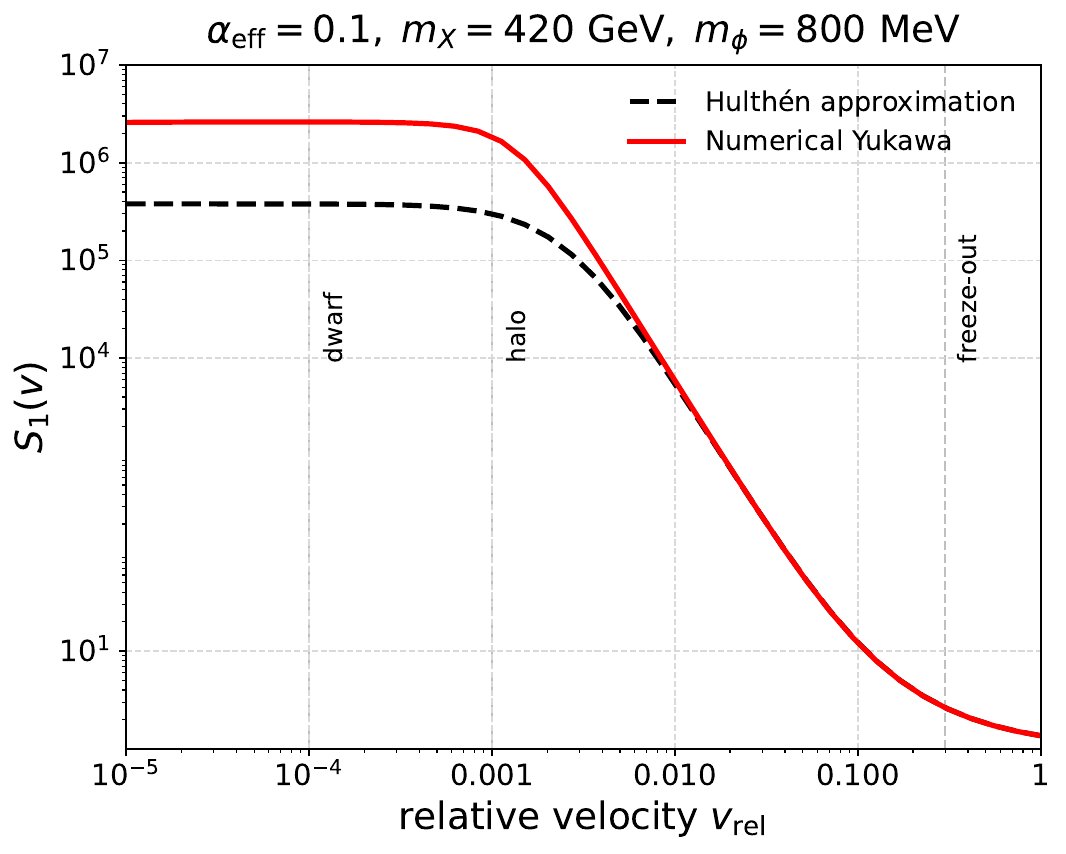}
\caption{Sommerfeld enhancement factor $S_1(v)$ obtained by numerically solving the Schr\"odinger equation with the exact Yukawa potential (solid red). The analytic Hulth\'en approximation (dashed black) is shown for comparison. We use $\alpha_{\rm eff}=0.1$, $m_X=420~\mathrm{GeV}$, and $m_\phi=800~\mathrm{MeV}$.}
   \label{fig:S1_yukawa}
\end{figure}
The resulting Sommerfeld enhancement factor is shown in Figure~\ref{fig:S1_yukawa}.
Motivated by the numerical Yukawa calculation, we adopt 
\begin{align}
m_X &= 420~\mathrm{GeV},\qquad m_\phi = 800~\mathrm{MeV},\nonumber \\ 
\alpha_{\mathrm{eff}} &= 0.1 \quad \left( \text{corresponding to } \mu \simeq 940~\mathrm{GeV} \text{ for } m_X = 420~\mathrm{GeV} \right)\nonumber
\end{align}
as our final benchmark point.
The corresponding benchmark parameters and phenomenological quantities are summarized in Table~\ref{tab:benchmark_num}. For comparison, the corresponding results obtained with the analytic Hulth\'en approximation are provided in Appendix~\ref{app:hulthen}.
\begin{table}[t] 
\centering 
\caption{Benchmark parameters and representative phenomenological quantities obtained using the numerical solution of the Schr\"odinger equation with the exact Yukawa potential.} 
\label{tab:benchmark_num} 
\begin{tabular}{lc} 
\hline\hline 
Quantity & Value \\ 
\hline 
DM mass $m_X$ & $420~\mathrm{GeV}$ \\ 
Mediator mass $m_\phi$ & $800~\mathrm{MeV}$ \\ 
Trilinear coupling $\mu$ & $940~\mathrm{GeV}$ \\ 
Coupling strength in the Yukawa potential $\alpha_{\rm eff}(m_X,\mu)$ & $0.1$ \\ 
Cutoff scale $\Lambda$ & $1~\mathrm{TeV}$ \\ 
\hline 
Freeze-out velocity $v_{\rm fo}$ & $0.3$ \\ 
Halo velocity $v_{\rm halo}$ & $10^{-3}$ \\ 
Dwarf velocity $v_{\rm dwarf}$ & $10^{-4}$ \\ 
\hline $S_1(v_{\rm fo})$ & $2.6$ \\ 
$S_1(v_{\rm halo})$ & $1.9\times10^{6}$ \\ 
$S_1(v_{\rm dwarf})$ & $2.6\times10^{6}$ \\ \hline 
$\langle \sigma v\rangle_{\rm fo}^{W^+W^-}$ & $4.9\times10^{-26}~\mathrm{cm}^3\,\mathrm{s}^{-1}$ \\ 
$\langle \sigma v\rangle_{\rm halo}$ & $4.0 \times 10^{-25}~\mathrm{cm}^3\,\mathrm{s}^{-1}$ \\ 
$\langle \sigma v\rangle_{\rm dwarf}$ & $5.5 \times 10^{-27}~\mathrm{cm}^3\,\mathrm{s}^{-1}$ \\ 
\hline\hline 
\end{tabular} \end{table}

For the benchmark point, we obtain $S_1(v_{\rm fo})\simeq 2.6$ at $v_{\rm fo}\simeq0.3$. This value is much smaller than the enhancement factor relevant for the Galactic halo and dwarf-galaxy environments and therefore does not qualitatively modify the relic-density prediction adopted in Ref.~\cite{Yamashita:2024krp}.
Using $\langle \sigma v \rangle_{\rm fo}^{W^{+}W^{-}} = 1.9 \times 10^{-26}~\mathrm{cm}^3\,\mathrm{s}^{-1}$,
which we obtained for DM pair annihilation
$XX \to W^{+}W^{-}$ at freeze-out without the enhancement factor, we estimate
\begin{align}
S_1(v_{\mathrm{fo}}) \langle \sigma v \rangle_{\rm fo}^{W^{+}W^{-}} \sim 4.9 \times 10^{-26}~\mathrm{cm}^3\,\mathrm{s}^{-1}.
\label{eq:freezeout_w_S}
\end{align}

We also obtain
\begin{align}
\langle \sigma v \rangle_{\rm halo}
&\sim S_1(v_{\mathrm{halo}})\,
\langle \sigma v \rangle_{\rm fo}^{W^{+}W^{-}}\,
\frac{v_{\rm halo}^2}{v_{\rm fo}^2}
\nonumber \\
&\sim 21\,
\langle \sigma v \rangle_{\rm fo}^{W^{+}W^{-}}
\sim 4.0 \times 10^{-25}~\mathrm{cm}^3\,\mathrm{s}^{-1}.
\end{align}
This result is consistent in order of magnitude with the estimate reported
by Totani,
$7.2 \times 10^{-25}~\mathrm{cm}^3\,\mathrm{s}^{-1}$,
for the $W^{+}W^{-}$ channel assuming an NFW$-\rho^2$ DM
morphology.
The inferred annihilation cross section is subject to astrophysical uncertainties associated with the choice of the DM halo profile.

To see constraints from dwarf galaxies, with
$S_1(v_{\mathrm{dwarf}}) \sim 2.6 \times 10^6$,
\begin{align}
\langle \sigma v \rangle_{\rm dwarf}
&\sim
S_1(v_{\mathrm{dwarf}})\,
\langle \sigma v \rangle_{\rm fo}^{W^{+}W^{-}}\,
\frac{v_{\rm dwarf}^2}{v_{\rm fo}^2}
\nonumber \\
&\sim 0.29\,
\langle \sigma v \rangle_{\rm fo}^{W^{+}W^{-}}
\sim 5.5 \times 10^{-27}~\mathrm{cm}^3\,\mathrm{s}^{-1}
< 10^{-25}~\mathrm{cm}^3\,\mathrm{s}^{-1},
\end{align}
which is consistent with the upper limits from dwarf spheroidal galaxies.
In the above estimate, we used $v_{\rm dwarf} \sim 10^{-4}$.
For smaller velocities, e.g., $v_{\rm dwarf} \sim 10^{-5}$,
the annihilation cross section is further suppressed.

At the epoch relevant for CMB constraints, the typical DM relative
velocity is extremely small.
After thermal freeze-out, the momenta of non-relativistic DM particles
redshift as $p \propto a^{-1}$, which implies
\[
v_{\rm CMB}
\sim
v_{\rm fo}\,\frac{T_{\rm CMB}}{T_{\rm fo}}
\sim
10^{-12},
\]
where we have used $v_{\rm fo}\sim0.3$, $T_{\rm fo}\sim20~\mathrm{GeV}$, and
$T_{\rm CMB}\sim0.25~\mathrm{eV}$~\cite{Baumann:2022mni}.
Here, $T_{\rm fo}$ and $T_{\rm CMB}$ denote the temperatures
at freeze-out and at the photon decoupling epoch, respectively.
Since the annihilation process is $p$-wave suppressed,
$\langle \sigma v \rangle \propto v^2$, and the Sommerfeld enhancement
saturates in the low-velocity regime, the annihilation rate at the photon
decoupling epoch is further suppressed compared to that in dwarf
spheroidal galaxies.
Consequently, the associated energy injection is negligible, and the
model safely satisfies the constraints from CMB observations.

We also note that the interpretation of the Galactic halo gamma-ray excess depends on the adopted diffuse emission model and other astrophysical assumptions. The benchmark considered here should therefore be regarded as an illustrative scenario motivated by the analysis of Ref.~\cite{Totani:2025fxx}.
We emphasize that the present analysis is not intended as a dedicated fit to the gamma-ray spectrum. Rather, our goal is to examine whether the DM mass and annihilation cross section preferred by the Galactic halo gamma-ray excess can be realized within the present model. A detailed likelihood or $\chi^2$ analysis is left for future work.

\section{Summary}
\label{sec:summary}

In this work, we have revisited a Higgs-portal spin-1 DM scenario
motivated by the halo-like gamma-ray excess reported by Totani.
The DM candidate is a dark photon associated with a gauged $U(1)_X$
symmetry, stabilized by a dark parity that forbids kinetic mixing
with the SM.
As a result, the leading interactions between the dark sector and the SM
arise from higher-dimensional operators.

When the short-distance annihilation of DM is governed by a
parity-violating dimension-six Higgs-portal operator,
$(H^\dagger H)X_{\mu\nu}\tilde X^{\mu\nu}$,
DM naturally evades stringent direct-detection constraints,
since the tree-level DM--nucleon scattering amplitude vanishes
in the zero-momentum-transfer limit,
and the annihilation is $p$-wave suppressed.
We have shown that thermal freeze-out driven by this operator
successfully reproduces the observed DM relic abundance
for a DM mass of around $400~\mathrm{GeV}$ with
an effective cutoff scale of $1~\mathrm{TeV}$.

This scenario predicts potentially observable indirect-detection
signals, in particular in the $W^{+}W^{-}$ channel.
The reported gamma-ray excess can be fitted by annihilation of DM
with a mass $m_X=420$~GeV for the $W^+W^-$ final state.

To explain this gamma-ray excess, we have introduced a light CP-even
scalar mediator $\phi$ that couples directly to the dark photon mass
operator, thereby enhancing the present-day annihilation rate in the
Galactic halo without qualitatively changing the thermal freeze-out picture.
The exchange of this scalar induces an attractive Yukawa potential
between non-relativistic DM particles,
giving rise to Sommerfeld enhancement of the annihilation rate.

Owing to the strong hierarchy of DM velocities in different
astrophysical environments, and for a benchmark choice of a DM
mass of $m_X = 420~\mathrm{GeV}$, a mediator mass of $m_\phi = 800~\mathrm{MeV}$,
and a dimensionful three-point $X$--$X$--$\phi$ coupling of
$\mu \simeq 940~\mathrm{GeV}$, the Sommerfeld enhancement becomes effective in
the Galactic halo ($v_\mathrm{halo} \sim 10^{-3}$), while it saturates in dwarf
spheroidal galaxies ($v_\mathrm{dwarf}  \sim 10^{-5}$--$10^{-4}$).
Combined with the intrinsic $p$-wave velocity suppression, this mechanism
selectively enhances the annihilation rate in the Milky Way halo, allowing
it to reach values of order
\[
\langle \sigma v \rangle_\mathrm{halo} 
\sim {\rm few}\times 10^{-25}~{\rm cm^3\,s^{-1}}
\]
which is compatible with the annihilation rate inferred from the gamma-ray excess reported by Totani,
while remaining well below current limits from dwarf spheroidal galaxies and the
CMB owing to the strong suppression of $p$-wave annihilation at very low
velocities, even with Sommerfeld enhancement.

\paragraph*{Acknowledgements}
We have used the package TikZ-Feynman~\cite{Ellis:2016jkw} to draw the Feynman diagrams.
This work was supported in part by JSPS KAKENHI Grant Number JP24K17040.

\paragraph*{Declaration of generative AI and AI-assisted technologies in the manuscript preparation process} 
During the preparation of this work, the author used ChatGPT (OpenAI), Microsoft Copilot, and Editage AI in order to facilitate idea development and language editing (ChatGPT and Microsoft Copilot), assist with drafting portions of Python code used in the numerical analysis (Microsoft Copilot), and conduct AI detection checks (Editage AI). After using these tools/services, the author reviewed and edited the content as needed and takes full responsibility for the content of the published article.

\appendix 
\section{Benchmark results obtained with the Hulth\'en approximation} 
\label{app:hulthen}
For reference, Table~\ref{tab:benchmark_hulthen} summarizes the benchmark parameters and phenomenological quantities obtained using the analytic approximation of Ref.~\cite{Cassel:2009wt}.
\begin{table}[t] 
\centering 
\caption{Benchmark parameters and representative phenomenological quantities obtained using the analytic Sommerfeld enhancement factor in Eq.~\eqref{eq:approx}, based on the Hulth\'en approximation of Ref.~\cite{Cassel:2009wt}.}
\label{tab:benchmark_hulthen}
\begin{tabular}{lc} 
\hline\hline 
Quantity & Value \\ 
\hline 
DM mass $m_X$ & $420~\mathrm{GeV}$ \\ 
Mediator mass $m_\phi$ & $400~\mathrm{MeV}/800~\mathrm{MeV}$ \\
Trilinear coupling $\mu$ & $1.3~\mathrm{TeV}/940~\mathrm{GeV}$ \\ 
Coupling strength in the Yukawa potential $\alpha_{\rm eff}(m_X, \mu)$ & $0.2/0.1$ \\ 
Cutoff scale $\Lambda$ & $1~\mathrm{TeV}$ \\ 
\hline Freeze-out velocity $v_{\rm fo}$ & $0.3$ \\ 
Halo velocity $v_{\rm halo}$ & $10^{-3}$ \\ 
Dwarf velocity $v_{\rm dwarf}$ & $10^{-4}$ \\ 
\hline $S_1(v_{\rm fo})$ & $6.1/2.6$ \\ 
$S_1(v_{\rm halo})$ & $1.3\times10^{7}/3.0\times10^{5}$ \\ 
$S_1(v_{\rm dwarf})$ & $3.1\times10^{7}/3.8\times10^{5}$ \\ 
\hline 
$\langle \sigma v\rangle_{\rm fo}^{W^+W^-}$ & $1.2\times10^{-25}~\mathrm{cm}^3\,\mathrm{s}^{-1}/4.9\times10^{-26}~\mathrm{cm}^3\,\mathrm{s}^{-1}$ \\ 
$\langle \sigma v\rangle_{\rm halo}$ & $2.7\times10^{-24}~\mathrm{cm}^3\,\mathrm{s}^{-1}/6.3\times10^{-26}~\mathrm{cm}^3\,\mathrm{s}^{-1}$ \\ 
$\langle \sigma v\rangle_{\rm dwarf}$ & $6.5\times10^{-26}~\mathrm{cm}^3\,\mathrm{s}^{-1}/8.0\times10^{-28}~\mathrm{cm}^3\,\mathrm{s}^{-1}$ \\ 
\hline\hline 
\end{tabular} 
\end{table}


\end{document}